\documentclass[12pt]{article}
\usepackage{epsf,cite,epsfig,psfig,float,amssymb,stmaryrd,latexsym}
\usepackage{graphics,psfrag}
\newcommand{\be}{\begin{equation}}  
\newcommand{\ee}{\end{equation}}  
\newcommand{\ba}{\begin{eqnarray}}  
\newcommand{\ea}{\end{eqnarray}}  
\newcommand{\nn}{\nonumber}
\newcommand{\vs}{\vspace{-.20cm}}
\newcommand{\vQ}{{\vec{Q}}}
\newcommand{\vsig}{\vec{\sigma}}

\newcommand{\vp}{\vec{p}}
\newcommand{\vk}{\vec{k}}
\newcommand{\vpp}{\vec{p}\,}

\newcommand{\hpp}{\hat{p}\,}
\newcommand{\wM}{\widetilde{M}}
\newcommand{\wG}{{\rm G}_d}
\newcommand{\newsection}[1]{\section{#1}\setcounter{equation}{0}}

\begin{document}
\thispagestyle{empty}
\hfill {\tiny \begin{tabular}{l}FZJ-IKP(TH)-2001-19
\\ FTUV-01-0918\\
IFIC-01-0918 \end{tabular}}

\vspace{2cm}

\begin{center}
{\bf {\Large Chiral dynamics and the reactions \\[0.3em]  
{\boldmath $pp \to d K^+ \bar{K}^0$} and       
{\boldmath $pp \to d \pi^+ \eta$}}}
\end{center}

\vspace{0.6cm}

\begin{center}
{\large E. Oset$^{a,b}$\footnote{email: oset@condor1.ific.uv.es},
Jos\'e A. Oller$^a$\footnote{email: j.a.oller@fz-juelich.de},
Ulf-G. Mei{\ss}ner$^a$\footnote{email: u.meissner@fz-juelich.de}}\\
\vspace{0.5cm}
{\small $^a$Forschungszentrum J\"ulich, Institut f\"ur Kernphysik (Theorie), 
D-52425 J\"ulich, Germany}\\
{\small $^b$Departamento de F\'{\i}sica Te\'orica and IFIC, Centro Mixto Universidad de 
Valencia-CSIC\\ Institutos de Investigaci\'on de Paterna, Aptad. 22085, 46071, 
Valencia, Spain}\\
\vskip 10pt

\end{center}

\vskip 20pt
\begin{abstract}
\noindent
We perform a study of the final state interactions of the $ K^+ \bar{K^0}$
and the $\bar{K^0}d$ systems in the reactions $pp \to d K^+ \bar{K^0}$ 
and $pp \to d \pi^+ \eta$. Since the two-meson system couples strongly to 
the $a_0(980)$ resonance, these reactions are expected to be an 
additional source of information about the controversial scalar sector. We also 
show that these reactions present peculiar features which can shed additional light 
 on the much debated meson-baryon scalar sector with strangeness $-1$. We deduce 
the general structure of the amplitudes close to 
the $d K^+\bar{K}^0$ threshold, allowing
for primary $ K^+ \bar{K^0}$ as well as $\pi^+\eta$ production with the two
mesons in relative S- or P-wave. The interactions of the mesons are accounted 
for by using  chiral unitary techniques,  which generate dynamically the
 $a_0(980)$ resonance, and the $\bar{K^0}d$ interaction is also 
taken into account. General formulae are  derived that allow to incorporate the 
final state interactions in these systems for any model of the production 
mechanism. We illustrate this approach by considering two specific production 
mechanisms based on three flavor meson-baryon chiral perturbation theory. It is 
demonstrated that in this scenario the $\bar{K^0}d$ interactions are very important 
and can change the cross section by as much as one order of magnitude. The amount of 
$\pi^+\eta$ versus $K^+ \bar{K^0}$ production is shown to depend critically on 
the primary  mixture of the two mechanisms, with large interference effects due 
to final state interactions. These effects are also shown to occur in the event 
distributions of invariant masses which are drastically modified by the final 
state interactions of the two-meson or the $\bar{K}d$ system.
\end{abstract}

\newpage


\section{Introduction}
The  reaction $pp \to d K^+ \bar{K^0}$ is presently the subject of 
experimental study by the ANKE collaboration at the Cooler Synchrotron 
COSY at J\"ulich with the aim (among others) of learning about the nature and properties 
of the $a_0(980)$ resonance \cite{anke}. The problem has attracted also 
the interest of theoretical groups \cite{cassing1,cassing2}(see furthermore the 
contributed papers in \cite{workshop}).
  The prospect of gaining novel information about the  $a_0(980)$ 
resonance, which might help to shed further light from the experimental side on 
the disputed nature of this resonance, is one of the attractive features of this 
reaction.  This controversy originates from  
the observation that there are several different models to deal with
the isospin 
$I=0 , \,1$ scalar sector, all of them reproducing the scattering data  to 
some extent, but with different  
conclusions with respect to the origin of the underlying dynamics. In particular, in  
refs. \cite{Torn,jaf,achasov,van,schech} these resonances are considered as preexisting 
ones (genuine quark model states),   
while in ref.  \cite{isgur2} they appear as meson--meson resonances
generated by a   
potential. In ref.  \cite{amsler} the advocated solution is that the  
$a_0(980)$ and the $f_0(980)$, the latter in the channel with $I=0$, are exotic resonances, 
that is, not simply $q\bar{q}$ states, while the   
preexisting $q\bar{q}$ scalar nonet should be heavier, around 1.4 GeV or so. 
Other interesting approaches to this problem are refs. 
\cite{pen,ulfi,markusin,stern,mex}, 
the relativistic quark model with an instanton-induced interaction of the Bonn 
group~\cite{Bonn}, 
the J\"ulich meson--exchange approach~\cite{Juel}, the Inverse Amplitude 
Method~\cite{ramonet} or some variants of it \cite{nieves}. It is notorious that 
opposite conclusions are obtained in refs. \cite{steele,narison} from the use of 
QCD sum rules. Regarding this controversy 
about the nature of 
the scalar 
resonances, the works of refs.\cite{npa,nsd,gamma,plb,marco,jpsi} have stressed 
the role of chiral symmetry and unitarity to understand the dynamics 
associated with the lowest lying scalar resonances (see also
\cite{ulfi}). As a result of the latter 
references  the  
lightest $0^{++}$ nonet is established to be of dynamical origin, i.e. made up 
of meson--meson  resonances, and is formed by the $\sigma(500)$,   
$\kappa$, $a_0(980)$ and a strong contribution to the physical $f_0(980)$. On the other  
hand, the preexisting scalar nonet would be made up by an octet around 1.4 GeV and a  
singlet contributing to the physical $f_0(980)$ resonance. Similar conclusions 
about the nature of the scalar resonances have been obtained in ref.\cite{steele} 
for the channels with isospin 0 and 1 within QCD sum rules. The previous set 
of works 
\cite{npa,nsd,gamma,plb,marco,jpsi} constitute a unique   
theoretical approach to the scalar sector capable to study all these reactions in 
an unified   way. This is achieved because all these processes are related by the 
use of an  effective theory description  that combines chiral
perturbation theory and unitarity constraints.  

It is important to remark, as already pointed out in ref.\cite{pen},
that it is mandatory to
study not only the experimental data concerning phase shifts and 
inelasticities related to meson-meson scattering but also production reactions where 
the interactions between the mesons manifest themselves through the final state 
interactions (FSI). This was indeed the subject of refs.\cite{gamma,plb,marco,jpsi} and
plays a key role in the present investigation.

  In the baryon sector the studies of \cite{weise,angels,ulfjose} based on the 
use of chiral Lagrangians also show
that the $\Lambda(1405)$ resonance is generated in a similar way. The fact that 
the $\Lambda(1405)$ is of dynamical origin was pointed out already many years 
ago \cite{dalitz}. More recently it has also been shown \cite{cornelius} that the 
$\Lambda(1670)$ and the $\Sigma(1620)$ are generated dynamically in the same chiral 
scheme of \cite{angels}. Still, in this sector more work is needed to
firmly establish these results.

As we will see, the reaction $pp \to d PQ$ (where $P,Q$ denote pseudoscalar  mesons)
offers novel possibilities with respect
to other reactions where the $a_0(980)$ resonance is produced  since
it is sensitive to {\em  both} the meson-meson and meson-baryon final
state interactions. In particular, we stress the importance of the
$\bar K d$ FSI, which has a very pronounced influence on observable
yields, invariant mass distributions or cross sections, mostly through
the interference with the meson-meson FSI (which gives rise to the
$a_0(980)$). This should provide 
extra information to test the implications of chiral symmetry on the nature of the 
low-lying resonances. 

The manuscript is organized as follows. In section~\ref{sec:brm}, we
discuss the basic reaction mechanisms for the process $pp \to d
PQ$, first in very general terms and
then we consider a specific model based on chiral symmetry for the
primary production of the meson pair. The final state interactions are
treated in section~\ref{sec:fsi}, separately for the meson-meson and
the meson-baryon systems. We stress in particular the role of the
anti-kaon-deuteron FSI. The results are presented and discussed in
section~\ref{sec:rd} and conclusions are drawn in
section~\ref{sec:con}. The appendix contains a detailed discussion on the
general structure of the process $pp \to d P Q$.

  
\section{Basic reaction mechanisms}
\label{sec:brm}

\subsection{General considerations}
The reaction measured in \cite{anke} is:
\be
\label{anke1}
pp\rightarrow d K^+ \bar{K}^0~.
\ee
We will study it theoretically in connection with the accompanying process:
\be
\label{anke2}
pp\rightarrow d\pi^+ \eta~,
\ee
since the dynamics of coupled channels, which we will use here, deals with both 
channels simultaneously. On the other hand, the energy of the ANKE experiment is 
fixed to 
$\sqrt{s}=2912.88$ MeV just about 45 MeV above the $dK^+\bar{K}^0$ threshold. 

The reaction (\ref{anke1}) forces the $K^+\bar{K}^0$ system to be in an $I=1$ state 
which, given the proximity of the $a_0(980)$ resonance, would have its rate of 
production and invariant mass distributions very much influenced by the tail of 
that resonance. The reaction (\ref{anke2}), which is also planned to be 
measured by the ANKE 
collaboration, could see the actual shape of the $a_0(980)$ resonance through 
the mass distribution of the $\pi^+\eta$ system.

The P-wave nature of the reaction \cite{cassing1,cassing2} is another peculiar 
feature that makes it different to other ones producing the 
$a_0(980)$\cite{a01,a02}. Indeed, due to total angular momentum and parity conservation 
as well as to the antisymmetry of the initial state, the two mesons cannot be 
simultaneously in intrinsic S-wave and in S-wave relative to the deuteron. Henceforth, 
we 
denote by $\ell$ the orbital angular momentum of the CM motion of the two 
pseudoscalars $PQ$ and the deuteron, and by $L$ the orbital angular momentum 
of the pseudoscalar mesons in their own CM frame, what we also call intrinsic angular 
momentum of the two mesons. With this notation the cases $\ell=1$, $L=0$ and 
$\ell=0$, $L=1$ 
are possible and then the initial state is forced to have $\ell_0=1,~3$ and $S=1$, 
with $\ell_0$ denoting the orbital angular momentum of the initial $pp$ state. 
These are the dominant contributions since they imply the lowest power, namely 1, 
of the small three-momenta of the deuteron or kaons in the transition amplitudes. 
Of course, the threshold of the reaction (\ref{anke2}) is much lower than the one of 
(\ref{anke1}) but due to the resonant nature of the interactions between the 
pseudoscalars, which only occurs when both $\pi^+\eta$ and $K^+\bar{K}^0$ are 
coupled together \cite{npa,nsd,ramonet}, we will consider the same structures 
as well for (\ref{anke2}). Even more, since the $\pi^+\eta$ system is expected 
to couple only very weakly to the P-waves then only the structure with $\ell=1$, $L=0$ 
is kept for $\pi^+\eta$. The fact that either $\ell=1$ or $L=1$ has its relevance 
since, close to threshold, one would expect  to have all particles 
in an S-wave. Thus, the $K^+\bar{K}^0$ system would be in an intrinsic S-wave 
state subject to the full strength of the $a_0(980)$ resonance tail. However, 
in the present case the contributions with $L=0$ or 1 are of the same order 
of magnitude. Furthermore, as commented above, if the $\pi^+\eta$ system does couple 
just weakly to the P-waves, as we will argue below that this is the expected behaviour, 
then the 
reaction mechanism establishes a distinction between the $K^+\bar{K}^0$ and $\pi^+\eta$ 
production processes which is novel with respect to other threshold production 
reactions.

The reaction mechanisms for $K^+\bar{K}^0$ and $\pi^+\eta$ production are complicated. 
Some approximate models have been used in refs. \cite{cassing1,cassing2} based on 
direct production of the $a_0(980)$ resonance. The $\pi^+\eta$ production channels 
are not
studied there although they could in principle be accounted for by using partial 
decay rates of the $a_0(980)$ into $K\bar{K}$ and $\pi\eta$ \cite{pdg}. However, the 
arguments given above about the unique prospects of this reaction indicate that the 
rates could be quite different in the present process than those determined 
by making use of standard Breit-Wigner parameterizations or related ones. Thus, 
substantial deviations from the rates observed in the present reaction 
would give further support to the  method of coupled channel generation of that 
resonance as the appropriate tool to deal with the light scalar resonances, versus 
static pictures that consider this resonance as an object with predetermined 
decay rates into different channels, and the final state interactions are just 
taken into account 
via Breit-Wigner like modifications of the resonance propagator. Indeed,  
inconsistencies in the treatment of the related $f_0(980)$ resonance 
as a pure Breit-Wigner or following a Flatt\'e formula have been recently 
pointed out in ref. \cite{turcos} in the reaction $\phi \rightarrow \gamma \pi^0\pi^0$.

\subsection{Dynamical model based on chiral symmetry}
\label{dmodel}
\begin{figure}[htb]
\centerline{\epsfig{file=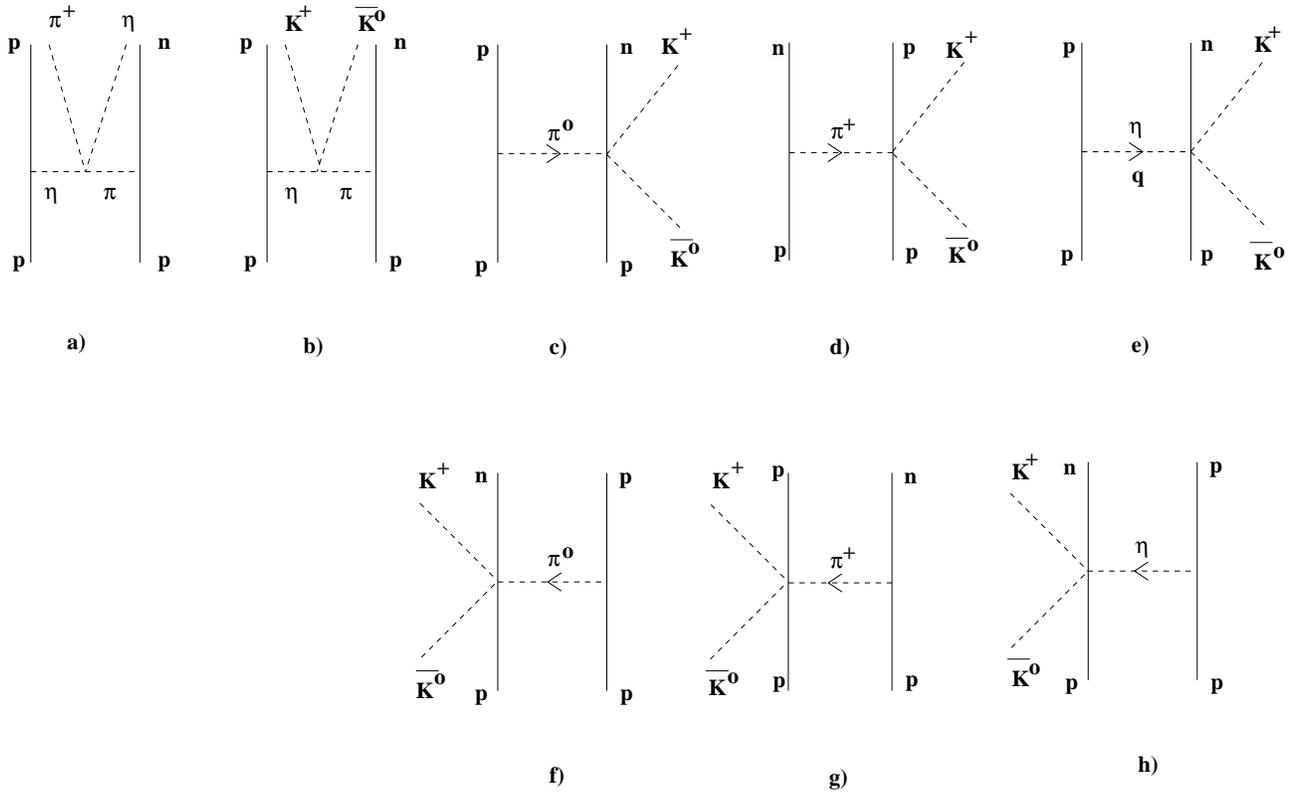,width=17cm}}
\vspace{1.3cm}
\caption[pilf]{\protect \small Chiral model for the primary production used to extract 
the structures given in eq.(\ref{mecs}).
\label{fig:mec1}}
\end{figure}

Since we are interested in stressing the role of the final state interactions of 
the $K\bar{K}$ and $\bar{K}d$ systems we refrain from searching for a complete
 model and 
simply parameterize the original structure of the amplitude close to the 
$dK\bar{K}$ threshold, the region of interest for the ANKE collaboration. This is 
done in Appendix A. Nevertheless the number of free parameters when allowing for 
the general structure deduced in Appendix A is too large to draw any definite 
conclusion. To overcome this difficulty we consider in the following a specific 
model derived from lowest order three flavor  chiral perturbation theory
(CHPT) \cite{gasser,ulf,pich} depicted in fig.\ref{fig:mec1}, 
giving rise to two definite structures which already illustrate the main 
points of our investigation, namely, the extreme importance of the final state 
interactions between the mesons and between the $\bar{K}^0$ and the deuteron. 
Then a variety of 
observables are evaluated in terms of the two parameters of the theory, 
up to a global normalization which will not be needed to evaluate ratios of cross 
sections and invariant mass distributions nor to compare with the future ANKE results. 
It is worth stressing that, given any model accounting 
for the primary production of the $dK^+\bar{K}^0$ and $d\pi^+\eta$ systems, the final 
state interactions can be taken into account following the general scheme presented 
in Appendix A. 

Let us come back to fig.\ref{fig:mec1}. Such diagrams involving the production of 
two mesons were evaluated in 
ref. \cite{ulftony} and have a similar structure, with some cancellations among them. 
In order to see this structure we consider diagrams \ref{fig:mec1}c) and 
\ref{fig:mec1}e). The two-baryon-three-meson (BBMMM) vertices are given by \cite{ulftony}
\be
\label{umu}
{\cal L}^{(B)}_1=\frac{1}{2}(D+F)\left(
\bar{n}\gamma^\mu \gamma_5 u_\mu^{(21)}p+\bar{p}\gamma^\mu 
\gamma_5 u_\mu^{(11)}p\right)~,
\ee
where $u_\mu$ is a SU(3) matrix containing the meson fields which is given explicitly 
in \cite{ulftony}. Furthermore, $D\simeq 3/4$ and $F\simeq 1/2$ are the canonical SU(3)
axial coupling constants. The relevant terms for our case are contained in the $u_\mu^{(21)}$ 
and $u_\mu^{(11)}$ 
matrix elements, and keeping in mind the non-relativistic reduction of $\gamma^\mu 
\gamma_5$,  are proportional to:
\ba
u_\mu^{(21)} &\rightarrow& \frac{3}{\sqrt{2}}\pi^0\left(\partial_\mu K^0 
 K^-  - \partial_\mu K^- K^0  \right)
+\sqrt{6}\,\partial_\mu\eta K^0 K^- \, ,\nn\\
u_\mu^{(11)} &\rightarrow& \partial_\mu \pi^+ K^0 K^-+
\pi^+\partial_\mu K^- K^0-2\pi^+ \partial_\mu K^0 K^-~.
\ea
Thus, diagram \ref{fig:mec1}c), with a non-relativistic reduction of $\gamma^\mu 
\gamma_5$ leads to a structure
\be
\label{p}
\frac{3}{\sqrt{2}}\vec{\sigma}^{(2)}(\vec{p}_{K^+}-\vec{p}_{\bar{K}^0})~,
\ee
which produces the $K\bar{K}$ system with a relative P-wave, while diagram 
\ref{fig:mec1}e) leads to a structure
\be
\label{newp}
\sqrt{6}\vec{\sigma}^{(2)}\vec{q}~,
\ee
where the superscript $1(2)$  applies to the the baryon line to the left(right) 
side of fig.\ref{fig:mec1}. In addition, the initial proton on the left(right) 
baryon line has three-momenta $\vp_1$($\vp_2$).

The global structure of the amplitude is obtained by considering also the 
$\vec{\sigma}^{(1)}\vec{q}$ 
vertex in the meson baryon vertex to the left side of the diagram. By taking
 $\vec{q}=\vec{p}_1-\vec{p}_d/2$, with $\vec{p}_1$, $\vec{p}_d$ the momenta of the 
initial proton and the deuteron, respectively (we have checked that consideration 
of the Fermi motion in the deuteron does not change the final structure), we find 
two types of terms:
\ba
\label{mecs}
&{\rm I)}& \vec{\sigma}^{(1)}(\vec{p}_1-\vec{p}_d/2)\,
\vec{\sigma}^{(2)}(\vec{p}_{K^+}-\vec{p}_{\bar{K}^0})-
\vec{\sigma}^{(1)}(\vec{p}_{K^+}-\vec{p}_{\bar{K}^0})\,
\vec{\sigma}^{(2)}(\vec{p}_2-\vec{p}_d/2)~, \nn\\
&{\rm II)}&\vec{\sigma}^{(1)}(\vec{p}_1-\vec{p}_d/2)\,
\vec{\sigma}^{(2)}(\vec{p}_1-\vec{p}_d/2)-
\vec{\sigma}^{(1)}(\vec{p}_2-\vec{p}_d/2)\,
\vec{\sigma}^{(2)}(\vec{p}_2-\vec{p}_d/2)~,
\ea 
which would come from the sum of the diagrams figs.\ref{fig:mec1}c) and 
\ref{fig:mec1}f), for eq.(\ref{mecs}.I), and diagrams figs.\ref{fig:mec1}e) and 
\ref{fig:mec1}h), for eq.(\ref{mecs}.II), after taking into account the 
isospin zero of the deuteron. Also the deuteron wave function 
appears with its value at the origin, $\phi(0)$, in coordinate space 
neglecting the range of the interaction, since the propagators of the 
pseudoscalar mesons in fig.\ref{fig:mec1} carry very high momentum transfers. 
In order to take into account the antisymmetry of the initial 
$pp$ state we must subtract to the previous expressions the same 
amplitudes exchanging the two initial protons. This means both spin and 
momentum but, since we have $S=1$ in the initial state, the wave function 
is spin symmetric and then it is enough to subtract the amplitudes of 
eq.(\ref{mecs}) exchanging $\vp_1 \leftrightarrow \vp_2=-\vp_1$. This leads, 
up to a global factor two, to the structures: 
\ba
\label{mecsf}
&{\rm I)}& \vec{\sigma}^{(1)}\vec{p}_1\, 
\vec{\sigma}^{(2)}(\vec{p}_{K^+}-\vec{p}_{\bar{K}^0})
+\vec{\sigma}^{(1)}(\vec{p}_{K^+}-\vec{p}_{\bar{K}^0})\,
\vec{\sigma}^{(2)}\vec{p}_1~, \nn\\
&{\rm II)}&-\vec{\sigma}^{(1)}\vec{p}_1\,\vec{\sigma}^{(2)}\vec{p}_d-
\vec{\sigma}^{(1)}\vec{p}_d\,\vec{\sigma}^{(2)} \vec{p}_1 ~.
\ea 

This calculation would not be complete to account for the $\pi$ exchange 
because the explicit use of the isospin deuteron wave function forces also the 
simultaneous consideration of diagrams figs.\ref{fig:mec1}d) and \ref{fig:mec1}g) with 
the 
exchange of a charged pion. The structure of these two latter diagrams is different 
than the structure eq.(\ref{mecs}.I) found for $\pi^0$ exchange, but it is 
easy to see that it is a combination of eqs.(\ref{mecs}I), (\ref{mecs}II) and 
after antisymmetrization with respect to the initial state leads again  to the 
structures of eqs.(\ref{mecsf}).

Should the $K\bar{K}$ system be in an intrinsic S-wave, $L=0$, we would have 
 eq.(\ref{mecsf}II) and the cross section contains the  factor $\vec{p}_d^{\;2}$, as 
correctly stated in ref. \cite{analysis}, which largely affects the shape of the 
$K\bar{K}$ invariant mass distribution.

As already mentioned, we will also consider $\pi^+\eta$ production. One interesting 
property of the $\pi\eta$ system reflected in chiral dynamics is that it does not 
couple in P-waves to lowest order in the chiral counting \cite{veronique}. It does 
not couple to vector mesons either \cite{derafael}. It can couple in higher orders 
but such effects are suppressed by more than one order of magnitude with respect to 
the dominant S-waves~\cite{veronique}.  This means that there are no terms of the 
type of diagram 
\ref{fig:mec1}c), d) or e) with the structure of eq.(\ref{p}), and in fact what one 
finds is that the matrix elements $u_\mu^{(21)}$ and $u_\mu^{(11)}$ of eq.(\ref{umu}) 
do not contain 
any $\pi\pi\eta$ or $\eta\eta\pi$ terms. However, terms of the type of diagram 
\ref{fig:mec1}a), with the $\pi\eta$ system in S-wave are allowed leading to a 
structure of the global amplitude of type II in eq.(\ref{mecsf}). The same 
comment applies to fig.\ref{fig:mec1}b) with respect to the $K^+\bar{K}^0$ state.

One still has to take into account that in the experiment only
unpolarized observables are measured. This 
implies an equal probability for the initial state of being in any of the three possible 
total spin projections. Taking the vector $\vQ$ to represent either 
$\vp_{K^+}-\vp_{\bar{K}^0}$ or $\vp_d$, any of the structures shown 
in eq.(\ref{mecsf}) can be written as 
${\cal A}\equiv \vsig^{(1)}\vp_1\,\vsig^{(2)}\vQ+\vsig^{(1)}\vQ\,\vsig^{(2)}\vp_1$.
It is straightforward to see that the matrix elements between states 
$|S\,S_3\rangle$ of well defined third component $S_3$ and total spin $S=1$ satisfy:
\be
\label{matrixelements}
\langle1\beta|{\cal A}|1\alpha\rangle=\eta_{\alpha\beta}
\frac{4\sqrt{\pi}}{\sqrt{3}}\,|\vp_1||\vQ|\,{\rm Y}_{1\,\alpha-\beta}(\hat{Q})~,
\ee
with $\hat{Q}$ the unit vector in the direction of $\vQ$, ${\rm Y}_{L m}(\theta,\phi)$ 
the usual  spherical harmonics and the 
matrix $\eta_{\alpha\beta}$ is given by:
\be
\label{cm}
\eta=\left(
\begin{array}{ccc}
+1 & -1 &  0 \\
+1 & -1 & +1 \\
 0 & -1 & +1  
\end{array}\right)~.
\ee
In this matrix the rows correspond to $\alpha=+1$, 0, $-1$, in order, and 
analogously 
for the columns. We have also taken $\vp_1$ parallel to the z-axis, that is, 
$\vp_1=(0,0,|\vp_1|)$.

The structures and couplings of the amplitudes discussed here could be evaluated 
in explicit microscopic models. Given the large energies involved there could be many 
competing mechanisms some of which would require information, like partial 
decays of resonances, which is not available at present. Given these limitations 
our choice provides a reasonable and simple starting point from the 
phenomenological side. It also illustrates the use of Appendix A in order to take 
care of the final state interactions in any other model.


\section{Final state interactions}
\label{sec:fsi}

In this section we first discuss the FSI due to the meson-meson interactions and 
then we will also consider the FSI from the $\bar{K}d$ channel. Afterwards, we sum up 
both contributions giving rise to our final renormalized  amplitudes.

\subsection{Meson-meson final state interactions}
The $K^+\bar{K}^0$ system in $I=1$ will interact strongly and couple to the $\pi^+\eta$ 
system. In \cite{npa} the input of the lowest order chiral Lagrangian was used as the 
kernel (potential) of the Bethe-Salpeter equation which produced exact unitarization 
in coupled channels. The extension in \cite{ramonet,nsd} to include effects from 
higher order Lagrangians, crossed channel dynamics, explicit exchange of 
genuine resonance states and scale independence did not modify the properties of the 
scalar sector found in \cite{npa} with only the lowest order Lagrangian and a cut-off 
of natural size, about 1 GeV, to regularize the loops. Since the divergences were 
only logarithmic the numerics are not changed when this cut-off is substituted by a 
regularization scale $\mu\sim 1$ GeV  and a subtraction constant that can be calculated in terms 
of the cut-off, see e.g.\cite{norbert,ulfjose,cornelius} \footnote{While the explicit calculations
have been done in dimensional regularization, this statement holds for any mass-independent
regularization scheme.}. Hence, given the 
simplicity and accuracy of ref. \cite{npa}, which is a limiting case of 
the more general formalism developed in 
\cite{nsd} (see e.g. \cite{ppnp,kyoto}),  we will use this approach in our 
present problem.

\begin{figure}[htb]
\centerline{\epsfig{file=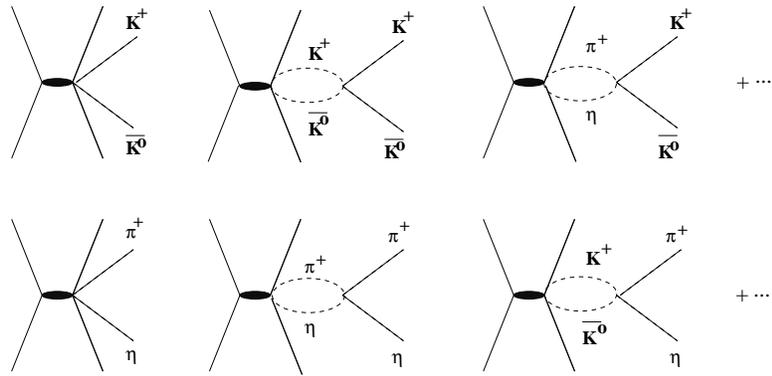,width=4.in}}
\vspace{0.3cm}
\caption[pilf]{\protect \small Diagrams relevant to take into account the 
meson-meson FSI. The $a_0(980)$ resonance is dynamically generated through 
the iteration of the meson-meson bubbles. This iteration is indicated in the 
figure by the ellipses.
\label{fig:mfsi}}
\end{figure}

Diagrammatically it means that in addition to the tree-level diagrams of 
fig.\ref{fig:mec1} we will have the diagrams of fig.\ref{fig:mfsi} which 
contribute to the $K^+\bar{K}^0$ production in the first line and to $\pi^+\eta$ 
production in the second one. By calling $G$ the loop function of the mesons, the 
sums  in fig.\ref{fig:mfsi} will dress the structure eq.(\ref{mecsf}.II) 
containing the $\vp_d$ vector corresponding to the case when the two mesons 
are in S-wave in their CM reference system, $L=0$. So we will have:
\be
\label{swavepd}
\begin{array}{cl}
\pi^+\eta: &f^S_{\pi\eta}|\vp_1||\vp_d|{\rm Y}_{1m}(\hat{p}_d)\rightarrow 
|\vp_1||\vp_d|{\rm Y}_{1m}(\hat{p}_d)\left(f^{S}_{\pi\eta}+
f^{S}_{\pi\eta}G_{\pi\eta}t_{\pi\eta\rightarrow \pi\eta}+
f^{S}_{K\bar{K}}G_{K\bar{K}}t_{K\bar{K}\rightarrow \pi\eta}\right)~,\\ & \\
K^+\bar{K}^0: & f^S_{K\bar{K}}|\vp_1||\vp_d|{\rm Y}_{1m}(\hat{p}_d)\rightarrow 
|\vp_1||\vp_d|{\rm Y}_{1m}(\hat{p}_d)\left( f^{S}_{K\bar{K}}+
f^{S}_{K\bar{K}}G_{K\bar{K}}t_{K\bar{K}\rightarrow K\bar{K}}+
f^{S}_{\pi\eta}G_{\pi\eta}t_{\pi\eta \rightarrow K\bar{K}}\right)~,
\end{array}
\ee
where the subscript $m$ corresponds to {\footnotesize$\alpha-\beta$} in the notation of 
eq.(\ref{matrixelements}), all the three-momenta refer to the CM frame of the 
$pp$ system and $f^S_{PQ}$ are the couplings of the two pseudoscalar meson systems 
with $L=0$. The latter are defined such that the minus global sign in 
eq.(\ref{mecsf}II) is reabsorbed by them. Eqs.(\ref{swavepd}) are more elegantly 
written in a 2$\times$2 matrix form as:
\be
\label{swave} 
|\vp_1||\vp_d|{\rm Y}_{1m}(\hat{p}_d)\left(1+t\,G\right)
\left\{
\begin{array}{c}
f^{S}_{\pi\eta}\\
f^{S}_{K\bar{K}}
\end{array}
\right\}~,
\ee
with $t_{ij}$ the S-wave transition matrix 
$K\bar{K},~\pi\eta\rightarrow K\bar{K},~\pi\eta$ in 
$I=1$ with index ``1'' for $\pi\eta$ and index ``2'' for $K\bar{K}$, and $G$ a diagonal 
matrix $G={\rm diag}(G_{\pi\eta},G_{K\bar{K}})$. Now taking into account the 
Bethe-Salpeter equation and the on-shell factorization of the potential in the loop 
functions involved in the strong interactions proved in \cite{npa}:
\be
\label{matrT}
t=t_2+t_2\cdot G \cdot t\; ; \quad t=\left[1-t_2\cdot G \right]^{-1}\!\!\cdot t_2~,
\ee
where $t_2$ contains the corresponding lowest order CHPT meson-meson scattering 
amplitudes \cite{npa}, we can write eq.(\ref{swave}) as:
\be
\label{swave2}
|\vp_1||\vp_d|{\rm Y}_{1m}(\hat{p}_d)\,{\cal D}^{-1}\left\{\begin{array}{c}
f^{S}_{\pi\eta}\\
f^{S}_{K\bar{K}}
\end{array}\right\}~,\ee
where ${\cal D}=\left[1-t_2\cdot G \right]$. Both $t$ and $G$ are functions of the invariant 
mass of the meson-meson system, $M_I^2$. The previous formalism to take into 
account FSI was originally employed in 
the calculation of $\gamma\gamma \rightarrow$meson-meson in ref. \cite{gamma} and 
later systematized in more general terms taking into account the analytic properties 
of the form factors in refs. \cite{jpsi,ffv}. In these last references it can be seen 
that the above results for taking care of  the FSI are exact when considering only the 
 right hand or unitarity cut. We will take the functions 
$f^{S}_{\pi\eta}$ and $f^{S}_{K\bar{K}}$ as constants since we are  concerned 
in the energy region available to the ANKE 
collaboration which reduces to just about 45 MeV above the $K^+\bar{K}^0$ 
threshold. Our approach of taking care only of the right hand cut corresponds 
to the expected dominance of the resonances $a_0(980)$ and $\Lambda(1405)$ 
which are very close to the $dK^+\bar{K}^0$ threshold. The chiral model of 
sec.~\ref{dmodel} gives rise to real couplings $f^{S}_{\pi\eta}$ and $f^{S}_{K\bar{K}}$, 
and so we will take them in the following. Nevertheless, the results of Appendix A 
can be equally applied to 
real or complex coupling functions  $f^{S}_{\pi\eta}$ and $f^{S}_{K\bar{K}}$.

\subsection{Anti-kaon-deuteron final state interactions}
Now we consider the FSI from the $\bar{K}d$ system. The interaction of the $K^+$ with 
the protons and neutrons is rather weak 
\cite{weise} and we will neglect it. However, this is not the case for the 
$\bar{K}^0n$ interactions which are very strong close to threshold due to 
the $\Lambda(1405)$ resonance below the $\bar{K}^0n$ 
threshold \cite{dalitz,weise,angels,ulfjose}. On the other hand what we need 
here is the $\bar{K}^0$ interaction with the deuteron that is quite strong close 
to the threshold due to extra reinforcement of the multiple scattering of the 
$\bar{K}$ in the deuteron as proved in multiple evaluations of this 
quantity using Faddeev equations \cite{toker,torres,bahaoui,barret,deloff}. A 
reanalysis of this quantity to the light of 
the new $\bar{K}N$ amplitudes generated in the chiral dynamical approach of 
\cite{angels} was done in \cite{kamalov}
 within the fixed scatterer approximation for 
the deuteron, which proves rather accurate comparing the results with those of the 
non-static calculation of \cite{deloff}. A sizeable $\bar{K}d$ scattering length of 
about 
$(-1.6+i1.9)$ fm is obtained in \cite{kamalov}. In order to take into account this 
extra interaction we first extrapolate the results of the $\bar{K}d$ scattering 
amplitude at threshold of \cite{kamalov} to the small finite $\bar{K}$ energies of the 
ANKE experiment \cite{anke}. For this purpose we rely upon the results for the 
$\bar{K}N$ scattering matrix found in \cite{angels}(fig.9) which show a drastic 
reduction of the real part of $t_{K^-p}+t_{K^-n}$ at $\sqrt{s}\simeq 1450$ MeV. Taking 
into account this fact plus the approximate good results of the impulse approximation 
for the imaginary part of the $\bar{K}d$ scattering length and the fast decline of 
the real part of the amplitudes from threshold on, suggest a quadratic interpolation 
between the results at threshold and the impulse approximation at 
$\sqrt{s}\simeq 1450$ MeV and beyond. Hence, for the general and illustrative 
purposes of the present chiral model to the primary production amplitudes of the 
$K\bar{K}$ and $\pi\eta$ channels, 
the following parameterization is expected to provide a sufficiently accurate 
description of the $\bar{K}d$ scattering amplitude:
\be
\label{td1}
\hbox{Re}~t_{\bar{K}d}(\wM_B)=a(\wM_B-\wM_{B0})^2+b(\wM_B-\wM_{B_0})+c
~,\nn
\ee
with
\ba
&&\begin{array}{ll}
a=4.32\cdot 10^{-4}\,\hbox{MeV}^{-3}~, & b=-1.55\cdot 10^{-2}\,\hbox{MeV}^{-2}~,\\
c=0.13\,\hbox{MeV}^{-1}~, & \wM_{B_0}=1432\,\hbox{MeV}~,
\end{array}
\ea
where $\wM_{B}^2=(p_{\bar{K}^0}-p_d/2)^2$ is the invariant mass of the $\bar{K}^0$ and 
the neutron in the deuteron. The imaginary parts are well 
approximated by the impulse approximation and are 
almost constants in the whole interval. We take finally:
\be
\label{td2}
\hbox{Im}~t_{\bar{K}d}(\wM_B)=b'(\wM_B-\wM_{B_0})+c'
\ee
with
\be
\begin{array}{ll}
b'=1.1\cdot 10^{-3}\,\hbox{MeV}^{-2}~, & c'=-1.5\cdot 10^{-1}\,\hbox{MeV}^{-1}~.
\end{array}
\ee
We apply the formulae (\ref{td1}) and (\ref{td2}) for $\wM_B<1.45$ GeV, and 
\be
t_{\bar{K}d}(\wM_B)=t_{\bar{K}d}(1.45~\hbox{GeV})~,
\ee
for $\wM_B>1.45$ GeV. We have also checked that substituting this limiting value
by a larger one only modifies mildly the resulting distributions.

\begin{figure}[htb]
\centerline{\epsfig{file=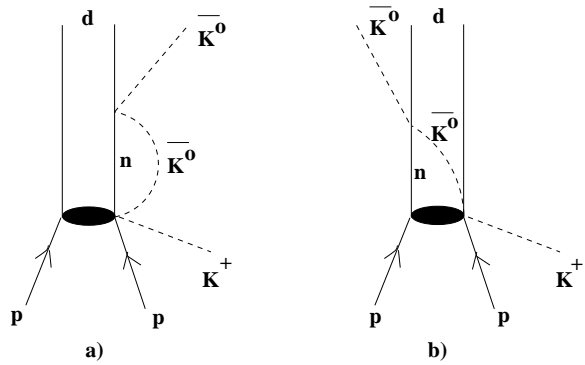,width=3.in}}
\vspace{0.3cm}
\caption[pilf]{\protect \small Diagrams to take into account the $\bar{K}^0d$ FSI.
\label{fig:kd}}
\end{figure}

The implementation of the FSI of $\bar{K}d$ requires to rewrite the amplitudes 
of eq.(\ref{matrixelements}), for $\vQ=\vp_{K^+}-\vp_{\bar{K}^0}$ and $\vQ=\vp_d$
 in the rest frame of the $\bar{K}d$. This is 
easily done by taking into account momentum conservation and the fact that 
$\vec{p}_{K^+}-\vec{p}_{\bar{K}^0}$ is a Galilean invariant. Given the 
small velocities involved in the $dK^+\bar{K}^0$ system we find it appropriate to 
just apply 
 Galilean transformations. Let us first consider the  intrinsic P-wave 
meson-meson contribution, eq.(\ref{mecsf}.I). In the following we denote by 
$\vp_\minuso=\vp_{K^+}-\vp_{\bar{K}^0}$, 
and hence this contribution involves ${\rm Y}_{1m}(\hat{p}_\minuso)$. Since we 
are considering Galilean invariance $\vp_\minuso=\vpp_\minuso'$, we then can 
rewrite $|\vp_\minuso|{\rm Y}_{1m}(\hat{p}_\minuso)$ as:

\ba
\label{pwave}
|\vp_\minuso|{\rm Y }_{1m}(\hat{p}_\minuso)=|\vp_\minuso|
{\rm Y}_{1m}(\hpp_\minuso')&=&
|\vpp_{K^+}'|{\rm Y}_{1m}(\hpp'_{K^+})-
|\vpp_{\bar{K}^0}'|{\rm Y}_{1m}(\hpp'_{\bar{K}^0})~,
\ea
where the primes stand for variables in the $\bar{K}^0d$ rest frame. The term 
${\rm Y}_{1m}(\hpp_{\bar{K}^0}')$ involves the P-wave contribution of the $\bar{K}^0$ 
and hence we neglect its modification in the present case of low $\bar{K}^0$ 
energies. The term ${\rm Y}_{1m}(\hpp_{K^+}')$ does not depend on 
$\vpp'_{\bar{K}^0}$. This implies that the deuteron and $\bar{K}^0$ are in a relative 
S-wave and can suffer the strong $\bar{K}d$ interaction.  The $\bar{K}d$ FSI are
 diagrammatically represented in fig.\ref{fig:kd}. 
and the corresponding terms are renormalized by changing them by:
\be
1+\wG \,t_{\bar{K}d}~,
\ee
where $\wG$ is the meson-deuteron loop function for the $\bar{K}N$ interaction. However, 
although the deuteron effects appear only as recoil corrections with respect to the 
nucleon meson-baryon $G_N$ loop function these effects can be around 
$m_K/M_d\simeq 25\%$ due to the large mass of the kaon, cf. fig.\ref{fig:kd}b).  
The function $G_N$ is given 
in \cite{ulfjose} using dispersion relations 
and a subtraction constant $a_{\bar{K}N}=-1.82$ as needed in the 
approach of \cite{cornelius}  to reproduce the low energy $\bar{K}N$ results of 
\cite{angels} using a cut-off to regularize the loops. Because of the already mentioned 
recoil effects and the large momentum transfer (through the shaded areas in 
fig.\ref{fig:kd}), we identify ${\rm G}_d=G_N$ but allow for a variation of 
 $\sim 30\%$ in $a_{\bar{K}N}$. Therefore, we present results for $a_{\bar{K}N}=-1.84$ 
and $-1.3$. Having said this, we find that eq.(\ref{pwave}) is renormalized by 
the FSI as:
\be
\label{pwavefsi}
|\vpp_{K^+}'|{\rm Y}_{1m}(\hpp'_{K^+})\left(1
+\wG \,t_{\bar{K}d}\right)-
|\vpp_{\bar{K}^0}'| {\rm Y}_{1m}(\hpp'_{\bar{K}^0})~.
\ee

Now, taking into account the following equalities:
\ba
\label{changeframe}
\vpp'_{K^+}&=&\vp_{K^+}\frac{M_d+2m_K}{M_d+m_K} ~, \nn\\
\vpp'_{\bar{K}^0}&=&\vp_{\bar{K}^0}+\vp_{K^+}\frac{m_K}{M_d+m_K}~,
\ea
it is then straightforward to rewrite eq.(\ref{pwavefsi}) as:
\be
\label{fsipwavef}
f_{K\bar{K}}^{P}|\vp_1|\,\left[|\vp_{K^+}| {\rm Y}_{1m}(\hat{p}_{K^+})\left(2
+\frac{M_d+2m_K}{M_d+m_K}\wG \,t_{\bar{K}d}\right)+|\vp_d| 
{\rm Y}_{1m}(\hat{p}_d)\right]~,
\ee
where we have multiplied the previous structure by the  $K^+\bar{K}^0$ 
P-wave coupling $f_{K\bar{K}}^{P}$ times $|\vp_1|$, since both factors appear in 
the original production process. In addition, $M_d=1875.61$ MeV is the deuteron mass and 
$m_K=495.7~\hbox{MeV}=(m_{K^+}+m_{\bar{K}^0})/2$, 
with $m_{K^+}=493.677$ MeV and $m_{\bar{K}^0}=497.672$ MeV the $K^+$ and $K^0$ masses, 
respectively.

We can follow similar steps to take into account the FSI of the $\bar{K}^0d$ 
to the CM meson-meson S-wave contribution, $L=0$, eq.(\ref{mecsf}.II). Considering 
the identity:
\be
\label{pwaved}
\vp_d=-\vpp'_{K^+}\frac{M_d}{M_d+2m_K}-\vpp'_{\bar{K}^0}~,
\ee
it then follows:
\be
\label{ypdpw}
|\vp_d|{\rm Y}_{1m}(\hat{p}_d)=-|\vp_{K^+}|{\rm Y}(\hpp'_{K^+})
\frac{M_d}{M_d+m_K}-|\vpp'_{\bar{K}^0}|{\rm Y}(\hpp'_{\bar{K}^0})~.
\ee
As discussed above the ${\rm Y}(\hpp'_{\bar{K}^0})$ term involves pure P-wave 
and it is not renormalized by the strong $\bar{K}^0d$ interaction. Its full 
contribution is then accounted for by eq.(\ref{swave2}). Hence only the 
term proportional to ${\rm Y}(\hpp'_{K^+})={\rm Y}(\hat{p}_{K^+})$, 
eq.(\ref{changeframe}), is renormalized due to the S-wave $\bar{K}^0d$ interaction as:
\be
\label{fsipwavepd}
-f_{K\bar{K}}^S||\vp_1||\vp_{K^+}|{\rm Y}(\hat{p}_{K^+})\frac{M_d}{M_d+m_K}\wG \,
t_{\bar{K}d}~,
\ee
multiplied by the corresponding coupling constant $f^S_{K\bar{K}}$ already 
introduced in eq.(\ref{swavepd}).

\subsection{Renormalized amplitudes}
Once we have taken into account the important FSI due to the resonant 
meson-meson and $\bar{K}^0d$ interactions, eqs.(\ref{swave2}), (\ref{fsipwavef}) 
and (\ref{fsipwavepd}), the renormalized $d\pi^+\eta$, $F_{\pi^+\eta}$, and 
$dK^+\bar{K}^0$, $F_{K^+\bar{K}^0}$, production amplitudes, 
corresponding to the transition between total spin third components 
$\alpha \rightarrow \beta$, read:
\ba
\label{Fs}
\frac{\sqrt{3}}{4\sqrt{\pi}}F_{\pi^+\eta}&=&
\eta_{\alpha\beta}|\vec{p}_1||\vec{p}_d|{\rm Y}_{1\,\alpha-\beta}(\hat{p}_d)
\bigg[[{\cal D}^{-1}(M_I^2)]_{11}f^S_{\pi\eta}+[{\cal D}^{-1}(M_I^2)]_{12}f^S_{K\bar{K}}\bigg]~,\nn\\
\frac{\sqrt{3}}{4\sqrt{\pi}}F_{K^+\bar{K}^0}&=&
\eta_{\alpha\beta}|\vec{p}_1||\vp_d|{\rm Y}_{1\,\alpha-\beta}(\hat{p}_d)
\bigg[[{\cal D}^{-1}(M_I^2)]_{21}f^S_{\pi\eta}+[{\cal D}^{-1}(M_I^2)]_{22}f^S_{K\bar{K}}
+f^P_{K\bar{K}}\bigg]\nn\\
&+&\eta_{\alpha\beta}|\vp_1||\vp_{K^+}|{\rm Y}_{1\,\alpha-\beta}
(\hat{p}_{K^+})\Bigg[
\frac{-M_d}{M_d+m_K}f_{K\bar{K}}^S \wG(\wM_B^2) \,t_{\bar{K}d}(\wM_B)\nn\\ 
&+&f_{K\bar{K}}^P 
\left(2+\frac{M_d+2m_K}{M_d+m_K} \wG(\wM_B^2) \,t_{\bar{K}d}(\wM_B)\right) \Bigg]~.
\ea

The double invariant mass distributions are obtained straightforwardly from 
the previous equations after summing over the final state polarizations and 
averaging over the initial ones. In this way one has: 
\ba
\label{dds}
\frac{d^{\,2}\sigma_{\pi^+\eta}}{dM_IdM_B }\!&=&\!\!16\pi{\cal C}\frac{|\vp_1|}{s^{3/2}}
\theta(1\!-\!|\cos\theta_{\pi^+}|)M_I M_B |\vp_d|^2\;\left|
[{\cal D}^{-1}(M_I^2)]_{11}f^S_{\pi\eta}+[{\cal D}^{-1}(M_I^2)]_{12}f^S_{K\bar{K}}\right|^2~,\nn\\
\frac{d^{\,2}\sigma_{K^+\bar{K}^0}}{dM_IdM_B }\!&=&\!\! 16\pi {\cal C} 
\frac{|\vp_1|}{s^{3/2}} \theta(1\!-\!|\cos\theta_{K^+}|)M_I M_B \Bigg\{ 
|\vp_d|^2\; \left| [{\cal D}^{-1}(M_I^2)]_{21}f^S_{\pi\eta}+[{\cal D}^{-1}(M_I^2)]_{22}
f^S_{K\bar{K}}+f^P_{K\bar{K}}\right|^2\nn\\
&+&|p_{K^+}|^2\;\left| \frac{-M_d}{M_d+m_K}f_{K\bar{K}}^S \wG(\wM_B^2) 
\,t_{\bar{K}d}(\wM_B)+f_{K\bar{K}}^P\bigg[ 2+\frac{M_d+2m_K}{M_d+m_K} \wG(\wM_B^2)
 \,t_{\bar{K}d}(\wM_B) \bigg]\right|^2
\nn\\
&+&2|\vp_d||\vp_{K^+}|\cos\theta_{K^+} \hbox{~Re}~\bigg[ 
\bigg([{\cal D}^{-1}(M_I^2)]_{21}f^S_{\pi\eta}+[{\cal D}^{-1}(M_I^2)]_{22}f^S_{K\bar{K}}+
f^P_{K\bar{K}}\bigg)^\star
\nn\\
&&\!\!\!\!\!\!\!\!\!\!\!\!\!
\times \left(\frac{-M_d}{M_d+m_K} f_{K\bar{K}}^S  \wG(\wM_B^2) \,t_{\bar{K}d}(\wM_B)+
f_{K\bar{K}}^P\bigg[2+\frac{M_d+2m_K}{M_d+m_K} \wG(\wM_B^2) \,t_{\bar{K}d}(\wM_B)
\bigg]\right)
\bigg]
\Bigg\}~,
\ea
with $s=(p_1+p_2)^2$ for the two protons in the initial state ($\sqrt{s}=2912.88$~MeV 
for the ANKE kinematics considered here), $M_B$ is the 
corresponding $\eta d$ or $\bar{K}^0d$ invariant 
mass and ${\cal C}$ is a normalization constant. The cosine of the angle 
between $\vp_{\pi^+}$($\vec{p}_{K^+}$) 
and $\vec{p}_d$, $\cos \theta_{\pi^+}(\cos\theta_{K^+})$, can be written in terms 
of $M_I$, $M_B$. By integrating with respect to $M_I$ or $M_B$ in eq.(\ref{dds}) we 
can obtain the invariant mass distributions with respect to $M_B$ and $M_I$, 
respectively.


\section{Results and discussion}
\label{sec:rd}

Apart from the absolute normalization of the amplitudes, our chiral 
model for the primary production depends on 
two free parameters, $\theta$ and $\phi$, such that
\begin{equation}
\label{parameterization}
f_{K\bar{K}}^S=\cos\theta ~,~ f^S_{\pi\eta}=\sin\theta \cos\phi~,~
f^P_{K\bar{K}}=\sin\theta \sin\phi ~.
\end{equation}  

\begin{figure}[htb]
\psfrag{KMI}{$d\sigma(K^+\bar{K}^0)/dM_I$}
\psfrag{MeV}{{\small M$_{\rm I}$ [MeV]}}
\centerline{\epsfig{file=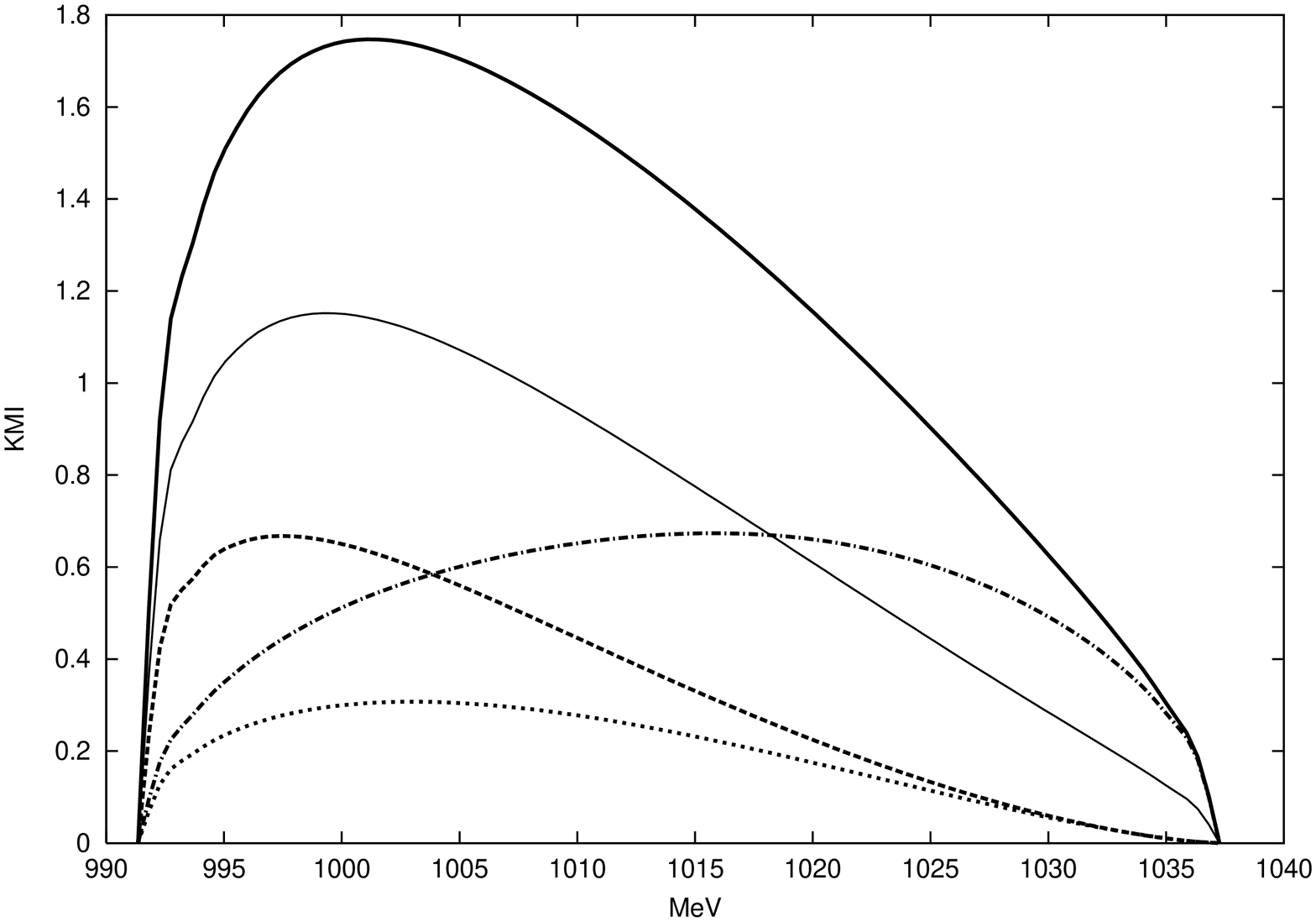,width=3.in,angle=-90}}
\vspace{0.3cm}
\caption[pilf]{\protect \small $d\sigma(K^+\bar{K}^0)/M_I$ for the whole range 
of available $M_I$ in the reaction $pp\rightarrow dK^+\bar{K}^0$ with 
$\sqrt{s}=2912.88$ MeV. The thick (thin) solid 
line is the full result with $a_{\bar{K}N}=-1.84 \, (-1.34)$.
The dashed line corresponds to including only 
meson-meson FSI, the dashed-dotted one includes only $\bar{K}^0d$ FSI and the 
dotted line includes no FSI with a $\vp_d^{\;2}$ factor for the modulus squared of the 
amplitude. 
\label{fig:comnnr}}
\end{figure}

 First, in order to show the relevance of the FSI, we take 
$\theta=\phi=0$, implying $f^S_{\pi\eta}=0$ and $f^{P}_{K\bar{K}}=0$. In 
fig.\ref{fig:comnnr} we display several curves  corresponding to 
$d\sigma_{K^+\bar{K}^0}/dM_I$ neglecting either the two 
considered FSI and including either one or two of them. In the rest of 
this section we take  $a_{\bar{K}N}=-1.84$ unless the contrary is explicitly stated. 
The distribution in the absence of any FSI (dotted line) peaks around $M_I=1003$ MeV. 
If the $K^+ \bar{K}^0$ interaction is switched on (dashed line) the strength is 
shifted considerably 
towards low invariant mass and the peak moves to about $M_I=997$ MeV. This 
is an obvious consequence of the presence of the $a_0(980)$ resonance around 980 MeV
and the $K^+\bar{K}^0$ distribution feels the tail of that resonance which increases 
the strength the closer one is to the resonance position, and hence to smaller 
values of the $K^+\bar{K}^0$ invariant mass. If one switches on only the $\bar{K}^0d$ 
FSI (dashed-dotted line) the distribution is rather broad and there is an 
accumulation of strength to higher values of the $M_I$. 
Finally, when all the interactions are considered (thick solid line)
the peak of the distribution moves back to lower masses around 1 GeV where the 
pure phase 
space peaks as well. The strength is furthermore increased by about a factor five 
due to 
the combined effects of both FSI. The effect of the $\bar{K}^0d$ interaction moving
the peak towards the center of the distribution reflects the fact that for these 
values of 
$M_I$ the $\bar{K}^0$ and the deuteron are at rest where ${\rm G}_d t_{\bar{K}d}$ 
has its  maximum. Indeed, in the 
extremes of the $M_I$ distribution either the kaons go together and the deuteron
goes opposite to them, or the deuteron is produced at rest and the two kaons go 
back to back. In both cases the $\bar{K}^0d$ invariant mass is relatively far from 
the $\bar{K}^0d$ threshold situation. In addition, we also include the full result 
for $a_{\bar{K}N}=-1.3$ presented by the thin solid line. As it is clear this 
variation in the value 
of $a_{\bar{K}N}$ mostly decreases the width of the distribution while the peak 
position is 
just slightly decreased by less than 2 MeV. Had we further reduced the value of 
$a_{\bar{K}N}$, instead of increasing it, the changes would have opposite sense.

\begin{figure}[htb]
\psfrag{KMB}{$d\sigma(K^+\bar{K}^0)/dM_B$}
\psfrag{MeV}{{\small M$_{\rm B}$ [MeV]}}
\centerline{\epsfig{file=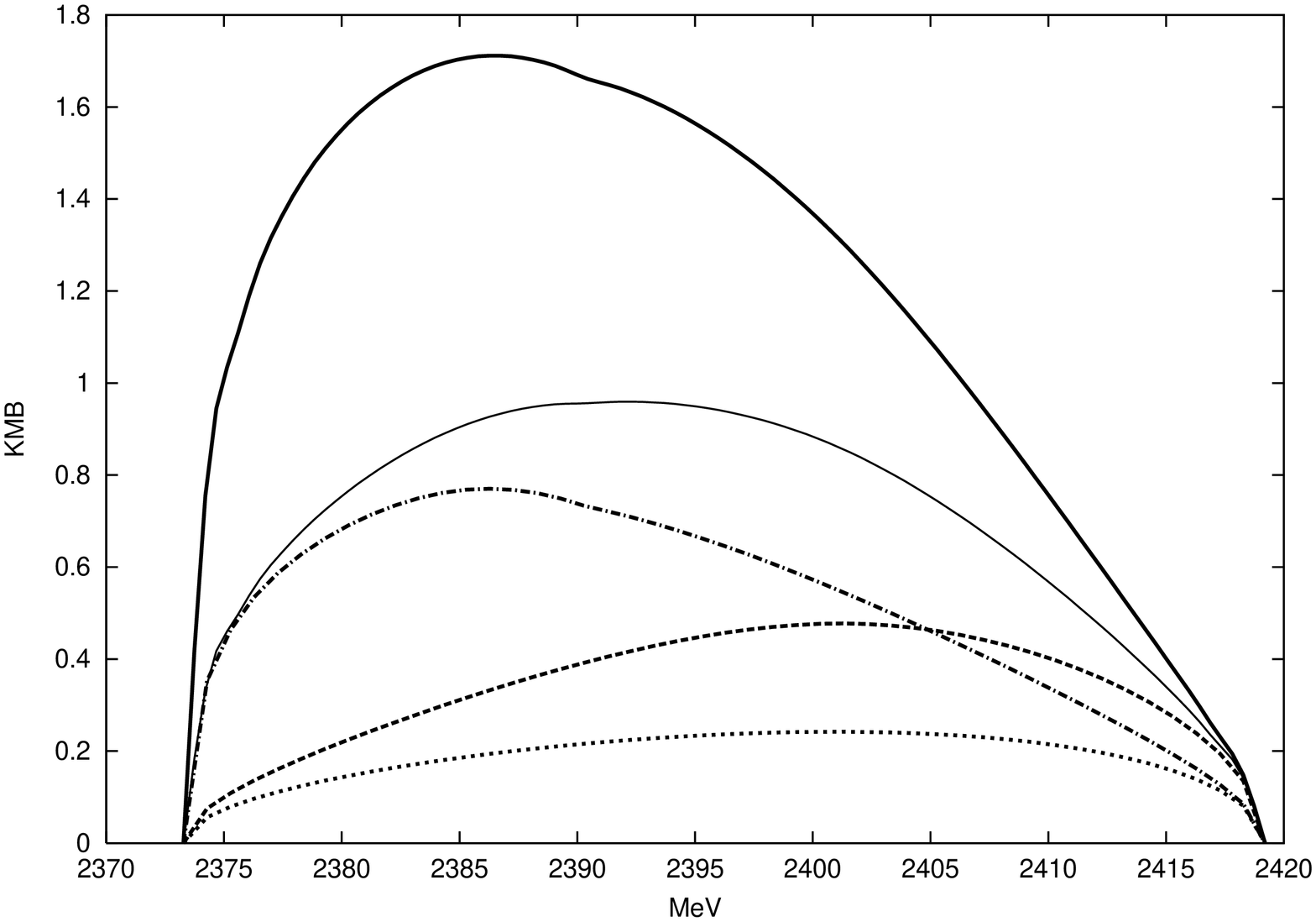,width=3.in,angle=-90}}
\vspace{0.3cm}
\caption[pilf]{\protect \small $d\sigma(K^+\bar{K}^0)/M_B$ for the whole range 
of available $M_B$ in the reaction $pp\rightarrow dK^+\bar{K}^0$  with 
$\sqrt{s}=2912.88$ MeV. The thick (thin) solid 
line is the full result with $a_{\bar{K}N}=-1.84\, (-1.34)$.
The dashed line corresponds to including only 
meson-meson FSI, the dashed-dotted one includes only $\bar{K}^0d$ FSI and the 
dotted line includes no FSI with a $\vp_d^{\;2}$ factor for the modulus squared of the 
amplitude. 
\label{fig:kdres}}
\end{figure}

  Similar changes in strength and shape can be seen in fig.\ref{fig:kdres} for the 
$d\sigma_{K^+\bar{K}^0}/dM_B$ invariant mass distribution where the notation for 
the curves is the same as for fig.\ref{fig:comnnr}. It is remarkable that 
the strong  $\bar{K}^0d$ interaction at threshold pushes the mass distribution 
towards lower $\bar{K}d$ invariant masses as a reflection of the presence of the 
$\Lambda(1405)$ resonance in the $\bar{K}^0n$ system, much as in the case of 
the $K^+\bar{K}^0$ distribution where the presence of the $a_0(980)$ resonance 
pushed the distribution 
towards low $K^+\bar{K}^0$ invariant masses. Here the effects of decreasing the 
modulus of 
$a_{\bar{K}N}$ are opposite to those in the $K^+\bar{K}^0$ mass distribution pushing 
the distribution to higher invariant masses.

\begin{figure}[htb]
\centerline{\epsfig{file=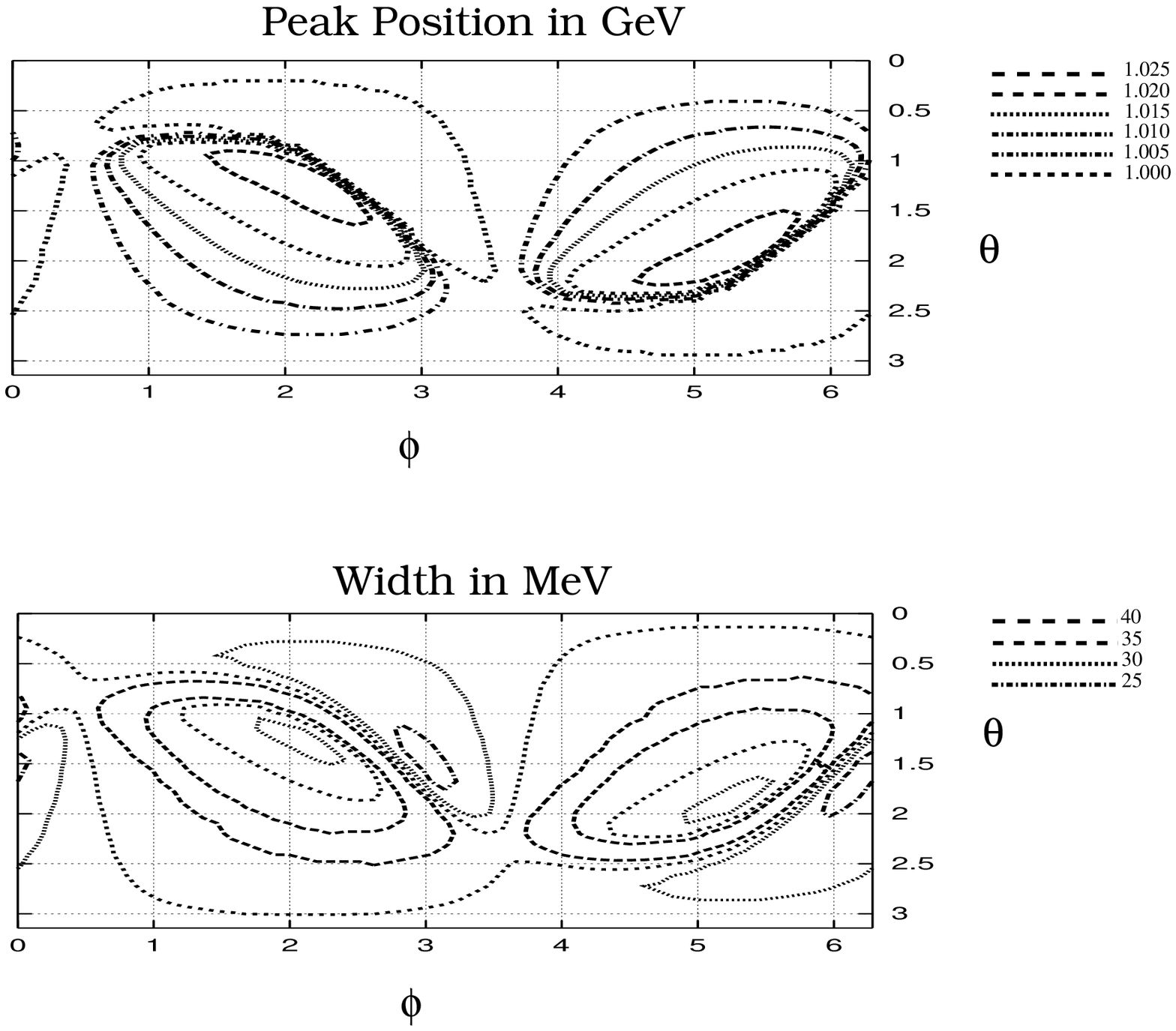,width=8.in}}
\vspace{0.3cm}
\caption[pilf]{\protect \small The upper panel shows the peak position of 
$d\sigma(K^+\bar{K}^0)/dM_I$ and the lower one the width of the same distribution for 
$a_{\bar{K}N}=-1.84$ as functions of $\theta$ and $\phi$ (in radians).
\label{fig:cp}}
\end{figure}

\begin{figure}[htb]
\centerline{\epsfig{file=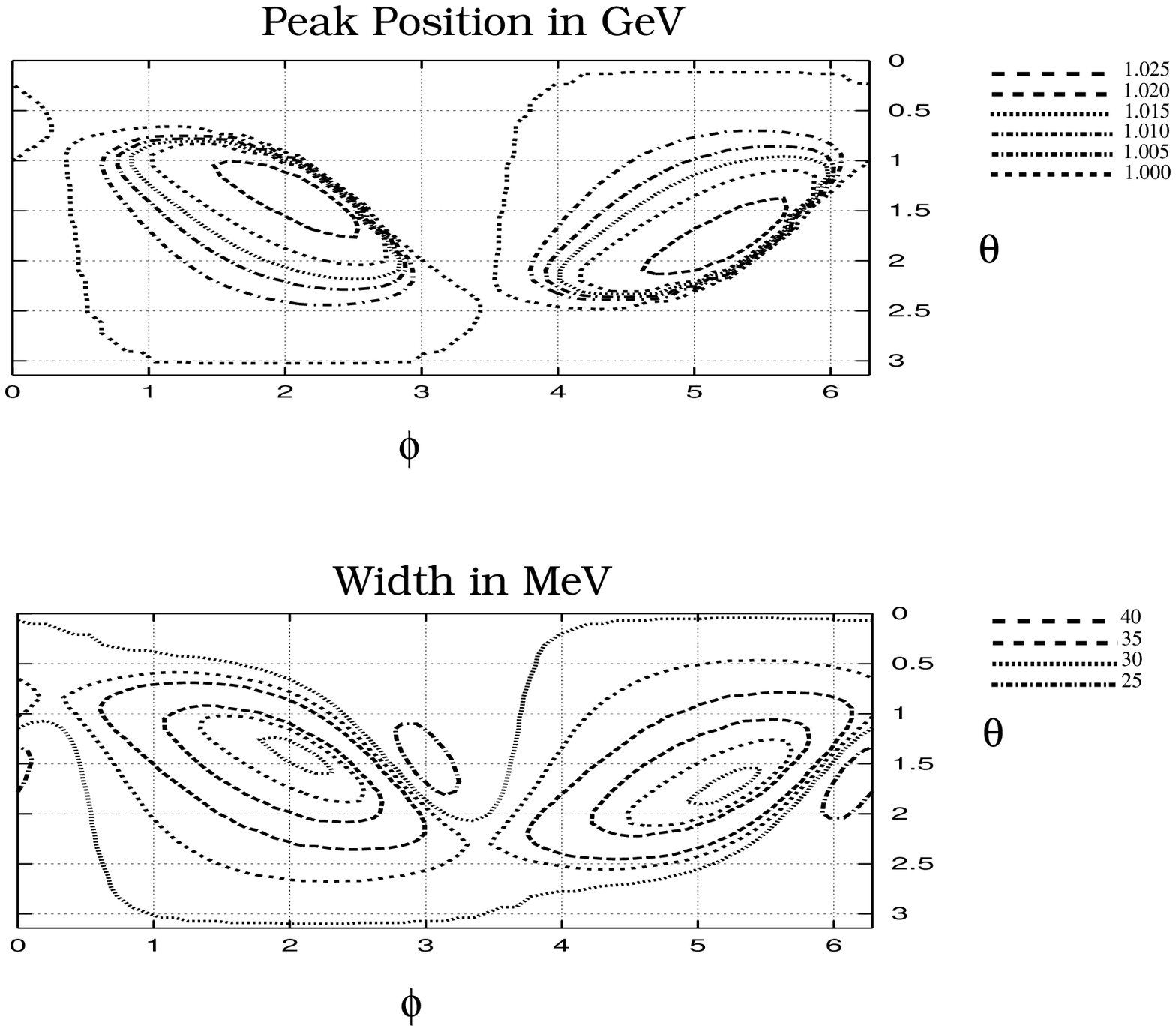,width=8.in}}
\vspace{0.3cm}
\caption[pilf]{\protect \small  The upper panel shows the peak position of 
$d\sigma(K^+\bar{K}^0)/dM_I$ and the lower one the width of the same distribution 
for $a_{\bar{K}N}=-1.3$ as functions of $\theta$ and $\phi$ (in radians).
\label{fig:cp2}}
\end{figure}

In fig.\ref{fig:cp} we show as a function of $\theta$ and $\phi$ the peak position 
of the $d\sigma_{K^+\bar{K}^0}/dM_I$ distribution and its width. The width is defined 
to be 
the difference between the values of $M_I$ for which the number of events is half 
of the maximum value. The same is shown  in fig.\ref{fig:cp2} for a different
value of the subtraction constant $a_{\bar{K}N}$, i.e.  
$a_{\bar{K}N}=-1.3$. The advantage of these figures is that one can approximately 
describe the shape of the distribution as a function of 
$\theta$ and $\phi$.

\begin{figure}[htb]
\centerline{\epsfig{file=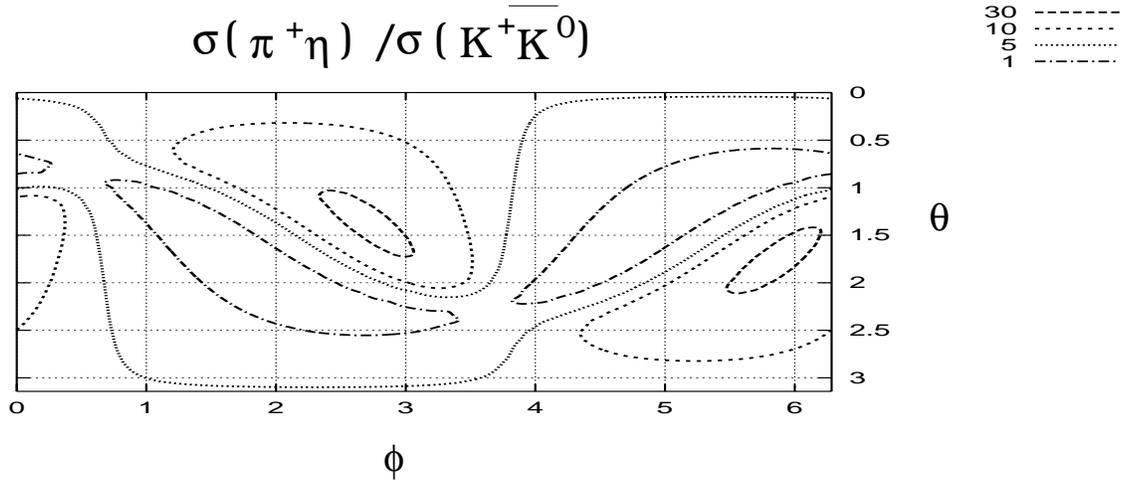,width=8.in}}
\vspace{0.3cm}
\caption[pilf]{\protect \small $\sigma(\pi^+\eta)/\sigma(K^+\bar{K}^0)$ as function 
of $\theta$ and $\phi$ (in radians).
\label{fig:ccpe}}
\end{figure}

  Next, we investigate the role of the $\theta$ and $\phi$ parameters  in the total 
production of $\pi^+\eta$ and $K^+\bar{K}^0$.  The  $\pi^+\eta$ production is mostly
done around the $a_0(980)$ resonance region. In fig.\ref{fig:ccpe} we show the ratio 
between the integrated 
$\pi^+\eta$ production cross section between $M_I=950$ MeV and the end of its 
phase space and the 
$K^+\bar{K}^0$ production cross section in all its available phase space. 
We can see in the figure that for most of the values of $\theta$ and $\phi$ 
the $\pi^+\eta$ production rate is substantially 
larger than that of $K^+\bar{K}^0$. It is interesting to point out that even in the 
case when there is no primary $\pi^+\eta$ production so that $f^S_{\pi\eta}=0$ 
($\theta=0$ and any $\phi$ or $\phi=\pi/2$, $3\pi/2$ and any $\theta$), the final state 
interactions starting from primary $K^+\bar{K}^0$ production can lead to a $\pi^+\eta$ 
cross section an order of magnitude bigger than that of the $K^+\bar{K}^0$. One also 
finds
interesting interference effects for some values of $\theta$ and $\phi$ that can 
reinforce the $\pi^+\eta$ production as compared to the $K^+\bar{K}^0$ as well as 
other situations when the $K^+\bar{K}^0$ is produced more copiously than the 
$\pi^+\eta$ channel. The former occurs for values of $\phi$ around $\pi$ and $2\pi$ 
and of $\theta$ around $\pi/2$ so that $f^S_{K\bar{K}}\simeq 0$ and $f^P_{K\bar{K}}
\simeq 0$ with $\sigma(\pi^+\eta)>30\, \sigma(K^+\bar{K}^0)$, while the latter tends 
to happen for a wide range of parameters leading to $K^+\bar{K}^0$ invariant 
mass distributions peaked at values of $M_I$ higher than 1010 MeV. This spectacular 
dependence of the 
ratio of $\pi^+\eta$ to $K^+\bar{K}^0$ production on the values of primary production 
weights $\theta$ and $\phi$ should obviously serve as a stimulation for the experimental 
measurement of the $\pi^+\eta$ production cross section.

\begin{figure}[htb]
\psfrag{PEI}{$d\sigma(\pi^+\eta)/dM_I$}
\psfrag{MeV}{{\small M$_{\rm I}$ [MeV]}}
\centerline{\epsfig{file=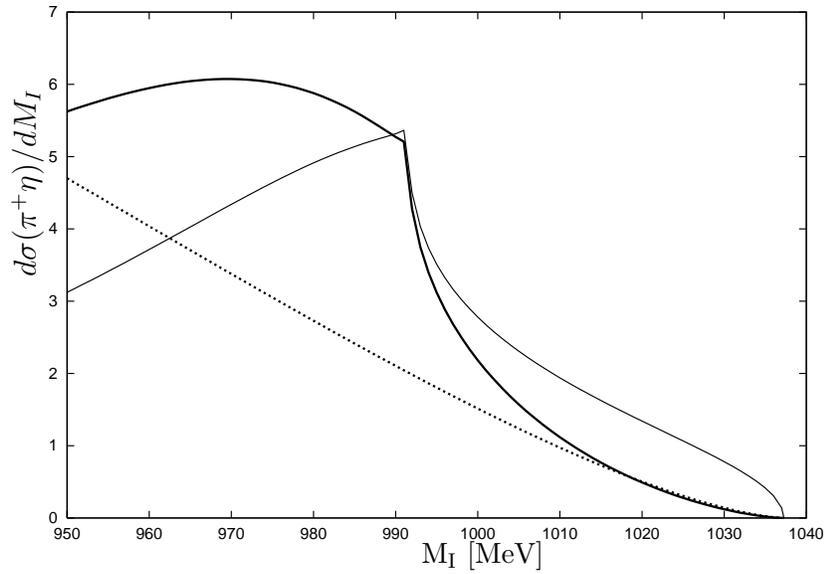,width=3.in,angle=-90}}
\vspace{0.3cm}
\caption[pilf]{\protect \small $d\sigma(\pi^+\eta)/dM_I$. Solid line, full result, 
dotted line does not include FSI, with a factor $\vp_d^{\;2}$ from the modulus 
squared of the amplitude. The thin solid line corresponds to the full result 
but divided by $\vp_d^{\;2}$ times $250^2$ MeV$^2$ (to normalize the
curve to the full result at the $\bar K K$ threshold).
\label{fig:pei}}
\end{figure}

  It is interesting to see also the shape of the $\pi^+\eta$ invariant mass 
distribution, which is independent of $a_{\bar{K}N}$. We show in fig.\ref{fig:pei} 
by the dashed-line that the normalized $d\sigma_{\pi^+\eta}/dM_I$ event distribution 
for $f^S_{\pi\eta}=0$ and $f^{P}_{K\bar{K}}=\,{\rm arbitrary}$ ($|f^{P}_{K\bar{K}}|\leq 
1$, see eq.(\ref{parameterization})) has no clear signal of the 
$a_0(980)$ resonance
around the values of $M_I=980$ MeV. This seems somewhat surprising since the 
coupled channel approach definitely generates the resonance and we have already 
observed the effects of its tail in the $K^+\bar{K}^0$ invariant mass distribution.
The lack of resonance structure is due to the P-wave character of the reaction 
and the appearance  of the $|\vp_d|^2$ factor in $|F_{\pi^+\eta}|^2$. This factor 
grows as the invariant mass decreases and distorts the $a_0(980)$ shape. In fact it is 
interesting to observe that if we divide  $d\sigma_{\pi^+\eta}/dM_I$  by  $|\vp_d|^2$,
 which is also shown in the figure by the thin solid line, the resonance 
shape appears 
with a width of around 40 MeV and with the peak at $990$ MeV. Noticing this fact is 
also important from the experimental point of view in order to extract properties of 
the $a_0(980)$ resonance in this reaction.

We can see that the distributions are rather dependent on the values of the 
 $\theta$ and $\phi$. This fact could be used to extract 
the optimal parameters from the data on $K^+\bar{K}^0$ distributions, assuming that 
a good fit is possible. 
Should this be the case, the theory would then predict absolute rates and 
mass distribution for the $\pi^+\eta$ production or other experimental yields,
 which would be a real prediction 
of the approach in spite of having started from two unknown parameters.

\section{Conclusions}
\label{sec:con}
In this paper we have performed a phenomenological study of the  
$pp \to d K^+ \bar{K^0}$ and $pp \to d \pi^+ \eta$ reactions
close to threshold presently studied by the ANKE collaboration at COSY.
We have emphasized the relevance of the final state interactions
which is quite important in the present case due to the proximity of the 
$K^+\bar{K}^0$ system to the $a_0(980)$ resonance and the $\bar{K}^0n$ system to 
the $\Lambda(1405)$ resonance. We found that the  consideration of these 
interactions has important consequences both in the shape and strength of the 
invariant mass distributions. We also studied the interaction of the two final 
states $K^+\bar{K^0}$ and $\pi^+\eta$ by means of a coupled channel chiral 
unitary approach which generates both the $a_0(980)$ and $\Lambda(1405)$ resonances. 
Given the 
freedom in the primary production amplitudes we parameterize them in terms of 
three types of structures involving $\ell=1$, $L=0$ and $\ell=0$, $L=1$ for 
the $K^+\bar{K}^0$ channel and the only 
allowed $\ell=1$, $L=1$ for $\pi^+\eta$ production. This left us with two 
independent parameters 
(up to a global normalization of one cross section) together with a subtraction 
constant with an expected uncertainty of about $25\%$. The sensitivity of the shapes 
of the $K^+\bar{K}^0$ and $\bar{K}^0d$ invariant mass distributions to those 
parameters was investigated 
anticipating that the measurements of these quantities could serve to fix them. This 
would allow us to make absolute predictions for $\pi^+\eta$ 
production due to the dynamics of coupled channels generated in the chiral unitary 
scheme followed here. Other quantities which might be measured could also be predicted 
in that case. Furthermore, we observed that the $\pi^+\eta$ production was dominated by 
the $a_0(980)$ resonance and that a clear signal for the relevance of the 
$\bar{K}^0d$ FSI would 
be the observation of a peak towards low $\bar{K}^0d$ invariant masses 
in the $d\sigma_{K^+\bar{K}^0}/dM_B$ differential cross section. On other hand, 
we have also pointed out 
that the $a_0(980)$ would not be clearly  visible in the 
data for $d\sigma_{\pi^+\eta}/dM_I$ because of the $|\vec{p}_d|^2$ factor due to 
the P-wave 
character of the reaction which distorts the shape of the resonance. Yet we  found 
that the shape of the resonance was 
regained by dividing $d\sigma_{\pi^+\eta}/dM_I$ by $|\vec{p}_d|^2$.

Finally, we have also provided general expressions to take into account the FSI 
derived in this paper for any other more specific model of the primary production 
mechanism. 

The study done here clearly shows how the measurements performed or planned with 
those reactions provide basic information on the strong interaction underlying the 
meson-meson and meson-baryon dynamics and should produce complementary and 
valuable information to the one obtained from other processes.

\vskip 20pt

\noindent
{\bf Acknowledgments}

\vskip 5pt
\noindent
One of us (E.O.) would like to acknowledge the hospitality of the Institut f\"ur
Kernphysik des 
Forschungszentrum J\"ulich where this work was carried out. We appreciate 
useful discussions with K. Kilian and  H. Str\"oher, V. Kobtev, M. B\"uscher, 
A. Sibirtsev and 
other members of the ANKE collaboration. This paper is partially supported 
by the DGICYT contract number BFM 2000-1326, the EU TMR network Eurodaphne, 
contract no. ERBFMRX-CT98-0169 and the Deutsche Forschungsgemeinschaft.

\appendix{}
\newsection{General structure of the process {\boldmath $pp\rightarrow d PQ$}}
\label{app:A}
\def\theequation{\Alph{section}.\arabic{equation}}
\setcounter{equation}{0}   

Let us denote by $\ell$ the relative orbital  angular momentum of the deuteron and 
the CM motion of the two pseudoscalar system $PQ$ and by $L$ the orbital angular 
momentum of the latter in their own CM frame. As discussed in sec.~\ref{sec:brm} 
close to the $dK^+\bar{K}^0$ threshold the leading contribution stems from 
$\ell=1$, $L=0$ and $\ell=0$, $L=1$. We denote by $\ell_0$ the orbital 
angular momentum of the two protons in the initial state and, as also discussed 
above, the only possibilities are $\ell_0=1,~3$. Finally, the symbol $S^d$ refers 
to the total spin of the deuteron, $S^d=1$, with its third component indicated 
by $S^d_3$. Analogously $S$ refers to the total spin of the $pp$ system, which 
is also fixed to be 1, and $S_3$ indicates its third component. 

Keeping only the components relevant for our reaction, we can consider the following 
angular momentum decomposition of the final $dPQ$ state as:
\ba
| d(S_3^d,\vp_d)P(\vk_1)Q(\vk_2)\rangle\!\!&\propto&\!\!\sum_{m,J} C(S_3^d~m|1~1~J)
\bigg({\rm Y}_{1m}(\hat{p}_d)^*|J,S_3+m;\ell=1,L=0\rangle\nn\\
&+&{\rm Y}_{1m}(\hat{k})^*|J,S_3+m;\ell=0,L=1\rangle\bigg)+...
\ea
where $\vk$ is the $PQ$-CM three-momentum of the pseudoscalar $P$, the symbol 
$C(m_1~m_2|j_1~j_2~J)$ is the Clebsch-Gordan coefficient for the composition of 
two angular momenta $j_1$ and $j_2$ to give the total one $J$ and the ellipses 
simply denote other terms of no interest here. It is 
worth noting once again that for the $\pi^+\eta$ system only the $\ell=1,~L=0$ 
component is relevant due to the absence of resonant interactions of this system 
with the deuteron. Performing an analogous decomposition for the initial $pp$ 
state we can write for the transition matrix element:
\ba
\label{tmat}
\langle d(S_3^d,\vp_d)P(\vk_1)Q(\vk_2) |T|p(\vp_1)p(\vp_2),S_3 \rangle=
\sum_{J,\ell_0}\!\!
&&\!\!\!\!\!\!C(S_3^d~S_3-S_3^d|1~1~J)~C(S_3~0|1~\ell_0~J)~{\rm Y}_{1\,0}(\hat{p}_1)^* 
\nn\\
&&\!\!\!\!\!\!\!\!\!\!\!\!\times\left({\rm Y}_{1\,S_3-S_3^d}(\hat{p}_d)T^{J\,PQ}_{10\ell_0}+
{\rm Y}_{1\,S_3-S_3^d}(\hat{k})T^{J\,PQ}_{01\ell_0}\right)
~,
\ea
where $\vp_1=(0,0,|\vp_1|)$ and because of 
the small 
velocities involved at around the $dK^+\bar{K}^0$ threshold we can also write 
${\rm Y}_{1m}(\hat{p}_\minuso)$ instead of ${\rm Y}_{1m}(\hat{k})$ with 
$\vp_\minuso$ defined in sec.~\ref{sec:fsi}. Note as well that we take the invariant 
matrix elements  $T^{J\,\pi\eta}_{01\ell_0}=0$, due to the absence of any resonant 
S-wave interaction between the $\pi^+\eta$ system and the deuteron. 
In order to take care of the final state interactions due to both the meson-meson 
and $\bar{K}d$ S-wave interactions one can proceed in a completely analogous way 
to that of sec.~\ref{sec:fsi}.
First we define the related quantitides $T^{J\,PQ}_{10\ell_0}=
|\vp_d|A^{J\,PQ}_{10\ell_0}$ and  $T^{J\,PQ}_{01\ell_0}=
|\vp_\minuso|A^{J\,PQ}_{01\ell_0}$ and second the $A_{\ell L
\ell_0}^{J\,PQ}$
can be taken, if desired, as constants.
In this way one has:
\ba
\label{gfsi}
F_{\pi^+\eta}&=&|\vp_d|{\rm Y}_{1\,S_3-S_3^d}(\hat{p}_d)\sum_{J,\ell_0}
{\rm Y}_{\ell_0\,0}(\hat{p}_1)^*\, C(S_3^d~S_3-S_3^d|1~1~J)~C(S_3~0|1~\ell_0~J)
\bigg([{\cal D}^{-1}(M_I^2)]_{11}A^{J\,\pi\eta}_{10\ell_0}\nn\\
&+&[{\cal D}^{-1}(M_I^2)]_{12}A^{J\,K\bar{K}}_{10\ell_0}\bigg)~,\nn\\
F_{K^+\bar{K}^0}&=&|\vp_d|{\rm Y}_{1\,S_3-S_3^d}(\hat{p}_d)\sum_{J,\ell_0} 
{\rm Y}_{\ell_0\,0}(\hat{p}_1)^*\, C(S_3^d~S_3-S_3^d|1~1~J)~C(S_3~0|1~\ell_0~J)
\bigg([{\cal D}^{-1}(M_I^2)]_{21}A^{J\,\pi\eta}_{10\ell_0} 
\nn\\
&+&[{\cal D}^{-1}(M_I^2)]_{22}A^{J\,K\bar{K}}_{10\ell_0}+A^{J\,K\bar{K}}_{01\ell_0}\bigg)+
|\vp_{K^+}|{\rm Y}_{1\,S_3-S_3^d}(\hat{p}_{K^+})\sum_{J,\ell_0}
{\rm Y}_{\ell_0\,0}(\hat{p}_1)^*\, C(S_3^d~S_3-S_3^d|1~1~J)\nn\\ 
&&\times C(S_3~0|1~\ell_0~J)\Bigg[\frac{-M_d}{M_d+m_K}\wG(\wM_B^2)\,t_{\bar{K}d}(\wM_B)
A^{J\,K\bar{K}}_{10\ell_0}
\\
&+&\left(2+\frac{M_d+2m_K}{M_d+m_K}\wG(\wM_B^2)\,
t_{\bar{K}d}(\wM_B)\right)A^{J\,K\bar{K}}_{01\ell_0}\Bigg]~.\nn
\ea
 
Once a model for the primary production mechanism of the $PQ$ systems is developed, 
the functions $T^{J\,PQ}_{\ell L \ell_0}$ can be determined and from them the FSI 
state interactions can be taken into account by eq.(\ref{gfsi}). We can do this 
exercise for our previous model  
by comparing eq.(\ref{tmat}) of the present appendix with eq.(\ref{matrixelements}), 
times the appropriate couplings constant
$f_{PQ}^{S}$ or $f_{K\bar{K}}^P$. Taking $\ell_0=1$, since our model 
only involves one power of $\vp_1$, we then have:

\be
\begin{array}{ll}
T^{0\,K\bar{K}}_{1\,0\,1}=-4\sqrt{3\pi}f_{K\bar{K}}^S|\vp_1||\vp_d|~, & 
T^{0\,K\bar{K}}_{0\,1\,1}=-4\sqrt{3\pi}f_{K\bar{K}}^P|\vp_1||\vp_\minuso|~,\\
&\\
T^{1\,K\bar{K}}_{1\, 0\, 1}=\frac{8\sqrt{\pi}}{\sqrt{3}}f_{K\bar{K}}^S|\vp_1||\vp_d|~,
& T^{1\,K\bar{K}}_{0\, 1\, 1}=\frac{8\sqrt{\pi}}{\sqrt{3}}f_{K\bar{K}}^P|\vp_1|
|\vp_\minuso|~,\\
& \\
T^{2\,K\bar{K}}_{1 0 1}=0~, & T^{2\,K\bar{K}}_{0 1 0}=0~,\\
& \\
T^{0\,\pi\eta}_{1\,0\,1}=- 4\sqrt{3\pi}f_{\pi\eta}^S |\vp_1||\vp_d|~, & 
T^{1\,\pi\eta}_{1\, 0\, 1}=\frac{8\sqrt{\pi}}{\sqrt{3}}f_{\pi\eta}^S|\vp_1||\vp_d|~,\\
& \\
T^{2\,\pi\eta}_{1 0 1}=0~. &
\end{array}
\ee

\newpage

\section*{Figures}

$\,$

\vspace{3cm}

\end{document}